\PassOptionsToPackage{table, prologue}{xcolor}

\documentclass[sigconf]{acmart}

\AtBeginDocument{%
  }

\setcopyright{acmlicensed}
\copyrightyear{2018}
\acmYear{2018}
\acmDOI{XXXXXXX.XXXXXXX}
\acmConference[Conference acronym 'XX]{Make sure to enter the correct
  conference title from your rights confirmation email}{June 03--05,
  2018}{Woodstock, NY}
\acmISBN{978-1-4503-XXXX-X/2018/06}

\usepackage{todonotes}
\usepackage{subcaption}
\usepackage{multirow}
\usepackage{makecell}
\usepackage{xcolor}
\usepackage{pgfplots}
\pgfplotsset{compat=1.17}

\definecolor{tsnec1}{RGB}{95,177,186}
\definecolor{tsnec2}{RGB}{209,93,103}
\definecolor{tsnec3}{RGB}{242,206,167}
\definecolor{tsnec4}{RGB}{209,130,98}
\definecolor{tsnec5}{RGB}{40,76,108}

\begin{document}

\newcommand{\model}{\textsc{LAMIA}}

%% The "title" command has an optional parameter,
%% allowing the author to define a "short title" to be used in page headers.
\title[\model{}]{
%TAILOR: Semantic Tokenization via  Learning Item Palette \\for Generative Recommendation
Learning Multi-Aspect Item Palette\paletteicon{}: A Semantic Tokenization Framework for Generative Recommendation}

\newcommand{\textclr}[2]{\texttt{\textcolor[HTML]{#1}{#2}}}
\newcommand{\textfmt}[4]{\textclr{#1}{#2:} \textclr{#3}{#4}}
\newcommand{\textttl}[1]{\textfmt{555555}{title}{555555}{#1}}
\newcommand{\textabs}[1]{\textfmt{555555}{abstract}{555555}{#1}}
\newcommand{\textcat}[1]{\textfmt{555555}{category}{555555}{#1}}
\newcommand{\texttsk}[1]{\textclr{847E35}{#1:}}
\newcommand{\textdto}[1]{\textclr{914444}{<LaP1>}}
\newcommand{\textdtt}[1]{\textclr{70A03D}{<LaP2>}}
\newcommand{\textdtr}[1]{\textclr{3D7C91}{<LaP3>}}
\newcommand{\textpho}[1]{\textclr{914444}{<LdP1>}}
\newcommand{\textpht}[1]{\textclr{70A03D}{<LdP2>}}
\newcommand{\textphr}[1]{\textclr{3D7C91}{<LdP3>}}
\newcommand{\mathdto}[1]{\mathbf{\textcolor[HTML]{957137}{#1}}}
\newcommand{\mathdtt}[1]{\mathbf{\textcolor[HTML]{2E5D86}{#1}}}

\newcommand{\news}[1]{\raisebox{-0.5ex}{\includegraphics[height=1em]{Assets/Icons/News#1.pdf}}}
\newcommand{\letter}[1]{\raisebox{-0.5ex}{\includegraphics[height=1em]{Assets/Icons/#1.pdf}}}

\newcommand{\palette}[2]{
    \begin{tikzpicture}[baseline=-0.6ex]
        \node[fill=#2, draw=black, text=black, rectangle, rounded corners=3pt, inner sep=2pt, draw, font=\scriptsize] {\texttt{#1}};
    \end{tikzpicture}
}

\definecolor{bglight}{RGB}{248, 248, 248}

\definecolor{bgdark}{RGB}{198, 248, 198}

% \definecolor{intrac}{RGB}{79,83,155}
% \definecolor{interc}{RGB}{149,110,71}
\definecolor{intrac}{HTML}{4285F4}
\definecolor{interc}{HTML}{0F9D58}

\newcommand{\intrac}[1]{\textcolor{intrac}{#1}}
\newcommand{\interc}[1]{\textcolor{interc}{#1}}

\definecolor{pc1}{rgb}{0.9098, 0.7647, 0.7647} % #E8C3C3: 232/255, 195/255, 195/255
\definecolor{pc2}{rgb}{0.8471, 0.9216, 0.7686} % #D8EBC4: 216/255, 235/255, 196/255
\definecolor{pc3}{rgb}{0.7412, 0.8667, 0.9098} % #BDDDE8: 189/255, 221/255, 232/255
\definecolor{pc4}{rgb}{0.7569, 0.5765, 0.5765} % #C19393: 193/255, 147/255, 147/255
\definecolor{pc5}{rgb}{0.9725, 0.9059, 0.9059} % #F8E7E7: 248/255, 231/255, 231/255

\newcommand{\paletteicon}{\raisebox{-0.5ex}{\includegraphics[height=1em]{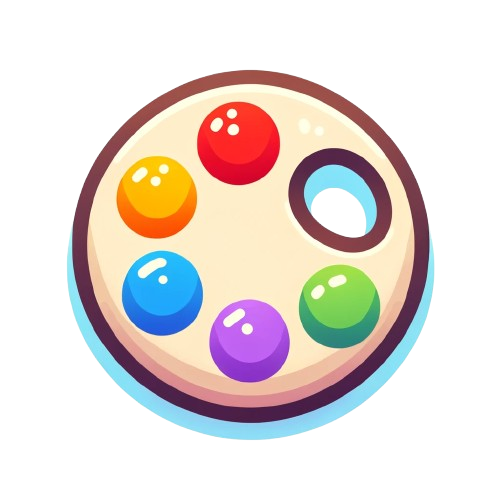}}}

%%
%% The "author" command and its associated commands are used to define
%% the authors and their affiliations.
%% Of note is the shared affiliation of the first two authors, and the
%% "authornote" and "authornotemark" commands
%% used to denote shared contribution to the research.
% \settopmatter{printacmref=false} 
\author{Qijiong Liu}
\affiliation{%
  \institution{The HK PolyU}
  \country{Hong Kong, China}
}
\email{liu@qijiong.work}

\author{Jieming Zhu}
\affiliation{%
  \institution{Huawei Noah's Ark Lab}
  \city{Shenzhen}
  \country{China}
}
\email{jiemingzhu@ieee.org}

\author{Zhaocheng Du}
\affiliation{
  \institution{Huawei Noah's Ark Lab}
  \city{Shenzhen}
  \country{China}
}
\email{zhaochengdu@huawei.com}

\author{Lu Fan}
\affiliation{%
  \institution{The HK PolyU}
  \country{Hong Kong, China}
}
\email{cslfan@comp.polyu.edu.hk}

\author{Zhou Zhao}
\affiliation{%
 \institution{Zhejiang University}
 \city{Hangzhou}
 \country{China}
}
\email{zhaozhou@zju.edu.cn}

\author{Xiao-Ming Wu}
\affiliation{%
  \institution{The HK PolyU}
  \country{Hong Kong, China}
}
\email{xiao-ming.wu@polyu.edu.hk}

%% By default, the full list of authors will be used in the page
%% headers. Often, this list is too long, and will overlap
%% other information printed in the page headers. This command allows
%% the author to define a more concise list
%% of authors' names for this purpose.
\renewcommand{\shortauthors}{Trovato and Tobin, et al.}

%%
%% The abstract is a short summary of the work to be presented in the
%% article.
\begin{abstract}
Traditional recommendation models often rely on unique item identifiers (IDs) to distinguish between items, which can hinder their ability to effectively leverage item content information and generalize to long-tailed or cold-start items. Recently, semantic tokenization has been proposed as a promising solution that aims to tokenize each item's semantic representation into a sequence of discrete tokens. These semantic tokens have become fundamental in training generative recommendation models. However, existing methods typically rely on RQ-VAE, a residual vector quantizer, for semantic tokenization. This reliance introduces several key limitations, including challenges in embedding extraction, hierarchical coarse-to-fine quantization, and training stability. To address these issues, we introduce \model{}, a novel approach for multi-aspect semantic tokenization. Unlike RQ-VAE, which uses a single embedding, \model{} learns an ``item palette''--a collection of \textbf{independent and semantically parallel embeddings} that capture multiple aspects of items. Additionally, \model{} enhances the semantic encoders through domain-specific tuning using text-based reconstruction tasks, resulting in more representative item palette embeddings. We have conducted extensive experiments to validate the effectiveness of the \model{} framework across various recommendation tasks and datasets. Our results demonstrate significant improvements in recommendation accuracy over existing methods. To facilitate reproducible research, we will release the source code, data, and configurations\footnote{\url{https://anonymous.4open.science/r/LAMIA/}}.

%However, existing generative recommendation methods typically rely on RQ-VAE, a residualvector quantizer, for semantic tokenization, which brings several key limitations on embedding extraction, hierarchical coarse-to-fine quantization, and training stability. To address these issues, in this paper, we introduce a novel approach, namely \model{}, for multi-asepct semantic tokenization. Instead of single embedding used in RQ-VAE, \model{} learns an ``item palette", which utilizes an matrix of independent embeddings to capture multiple aspects of items. \model{} also conducts domain-specific tuning of semantic encoders by using text-based reconstruction tasks, yielding morerepresentative item palette embeddings. Extensive experiments have been conducted to validate the effectiveness of our \model{} framework across various recommendation tasks and datasets. We will release the source code and configurations for reproducible research.
\end{abstract}

\ccsdesc[500]{Information systems~Recommender systems}

%%
%% Keywords. The author(s) should pick words that accurately describe
%% the work being presented. Separate the keywords with commas.
\keywords{Semantic Tokenization, Generative Recommendation, Large Recommendation Model}

\received{20 February 2007}
\received[revised]{12 March 2009}
\received[accepted]{5 June 2009}

%%
%% This command processes the author and affiliation and title
%% information and builds the first part of the formatted document.
\maketitle

\section{Introduction}

Sequential recommendation has been widely used in online applications, such as e-commerce websites, advertising networks, streaming services, and social media, to deliver personalized, timely recommendations tailored to users' interests. However, traditional sequential recommendation models often rely on unique item identifiers (IDs) to represent items~\cite{IDvsModality,MMRecSurvey}, which presents several key limitations. First, ID-based item representation can suffer from overfitting due to the typically sparse and imbalanced nature of the training data. Second, it fails to adequately leverage item content information, which is crucial for improving recommendations for long-tailed and cold-start items.

\begin{figure}[t]
    \centering
    \includegraphics[width=\linewidth]{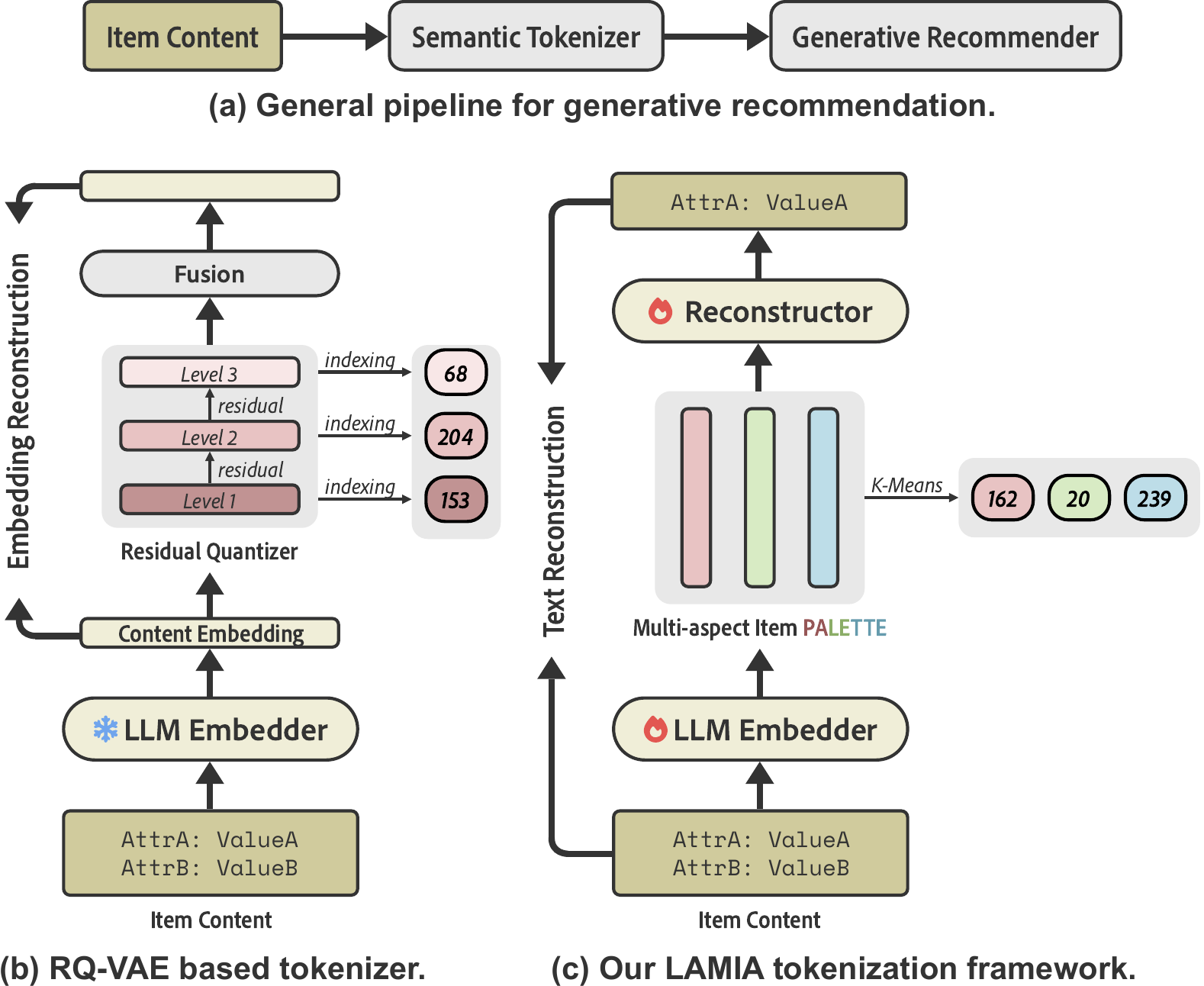}
    \caption{Illustrations of the generative recommendation paradigm, previous RQ-VAE based semantic tokenization approach, and our proposed \model{} framework.}
    \label{fig:comparison}
\end{figure}

To address these limitations, semantic tokenization has recently emerged as a promising solution and quickly gained attention within the community~\cite{semanticids,tiger,uist,CoST}. Instead of representing each item with a unique ID embedding, semantic tokenization encodes each item's semantic representation into a compact sequence of \textit{discrete tokens}, namely \textit{semantic identifier or semantic tokens}. These tokens can be shared across items, allowing the similarity between two items to be roughly estimated by the Hamming distance between their token sequences. The semantic identifiers can be obtained without the need for training on downstream recommendation tasks. As shown in Figure~\ref{fig:comparison}(a), once these semantic identifiers are generated, they can be utilized for generative recommendation \cite{tiger}. Consequently, semantic tokenization effectively addresses the aforementioned challenges and has demonstrated significant potential \cite{tiger,semanticids}.

\begin{figure*}[t]
    \centering
    \includegraphics[width=.7\linewidth]{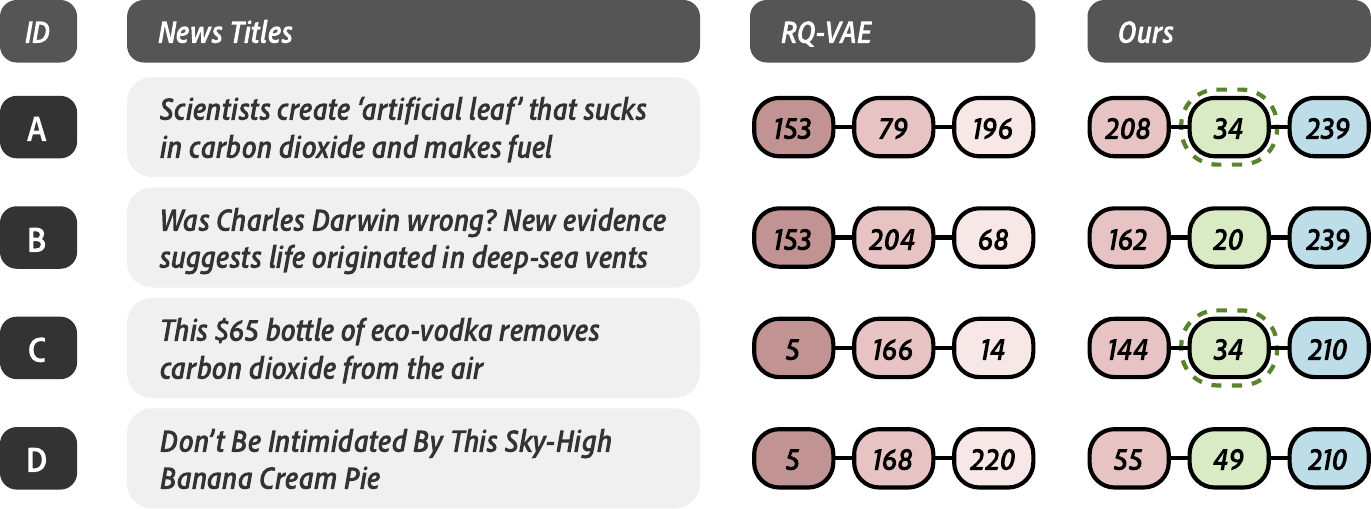}
    \caption{A case study on a news recommendation dataset, i.e., MIND. Under RQ-VAE, \news{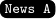} (\letter{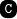}) is classified alongside \news{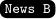} (\letter{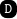}) in the science (food) category, failing to capture the shared environmental theme of \letter{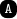} and \letter{C}. In contrast, \model{} captures news semantics from multiple aspects, correctly grouping \letter{A} and \letter{C} in the second aspect.}
    \label{fig:case}
\end{figure*}

Most existing methods~\cite{tiger,letter} utilize RQ-VAE \cite{rq-vae}, a residual vector quantizer, for semantic tokenization. This approach uses a differentiable hierarchical clustering mechanism to convert item representations into discrete semantic identifiers. As illustrated in Figure~\ref{fig:comparison}(b), the initial layer captures broad, coarse-grained semantic information, while subsequent layers refine this by quantizing residual vectors to capture fine-grained details.

However, RQ-VAE presents several critical limitations. Primarily, it focuses on capturing only the dominant semantic aspect of an item, determined by the clustering in the first layer. The subsequent layers merely enhance the details of this primary aspect. While this approach can be effective for certain datasets, it falls short in representing the complex, multifaceted nature of items, such as those involving diverse functionalities or varied topics. Figure~\ref{fig:case} illustrates an example. \news{A}, centered on science and environment, is classified to the science category \palette{153}{pc4}, whereas~\news{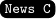}, concerning food and environment, is assigned to the food category \palette{5}{pc4}.
Clearly, this approach fails to reveal that both articles also belong to the environment category, resulting in significant information loss during the tokenization process. Secondly, the training of RQ-VAE is quite sensitive and prone to code collapse issues~\cite{CodeCollapse3,CodeCollapse2,CodeCollapse}.
Additionally, existing methods often use embeddings from pretrained encoders (e.g., LLMs) directly for quantization. This approach fails to capture domain-specific knowledge and data distribution, resulting in the generation of ineffective semantic identifiers by RQ-VAE.

In this work, we introduce a novel framework for semantic tokenization via learning multi-aspect item palette (\model{}). This framework, as illustrated in Figure~\ref{fig:comparison}(c), is orthogonal to RQ-VAE and is specifically designed to address the limitations of methods based on RQ-VAE.
Instead of quantizing multi-level residual vectors, we propose learning an ``item palette,'' a collection of vectors for a single item designed to capture its multi-aspect information. This structure is trained using a text-based reconstruction task, which minimizes information loss—unlike RQ-VAE, which relies on an embedding-based reconstruction task. Additionally, a contrastive learning task is employed to ensure the learning of independent, multi-aspect information for each item. The learned multi-aspect item representations are then quantized to generate multiple semantic codes using simple clustering algorithms, effectively bypassing the training challenges associated with RQ-VAE. As illustrated in Figure~\ref{fig:case}, the news pairs (\letter{A}, \letter{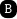}) and (\letter{C}, \letter{D}), categorized respectively under science and food themes, share an identical third-aspect code. Additionally, the environmental theme shared by \news{A} and \letter{C} is represented in the second aspect, aligning them both within category \palette{34}{pc2}.

In summary, our \model{} framework offers these advantages:

% \begin{itemize}
    % \item 
    \textbf{Parallel semantic identifier:} We present a novel approach for learning a multi-aspect palette -- a set of mutually exclusive, equally weighted, and semantically parallel embeddings that collectively capture distinct facets of item content. 

    % \item \textbf{Simple, training-free quantization process:} 
     % The learned multi-aspect representations can be quantized using simple, training-free clustering algorithms, thereby avoiding the training challenges inherent in RQ-VAE.

    % \item 
    \textbf{Text-based reconstruction approach:} Standard RQ-VAE methods rely on embedding-based reconstruction, which depends on high-quality embeddings from pretrained LLMs and may suffer from information loss due to data distribution shifts. In contrast, our domain-adaptive strategy uses text-level reconstruction with domain-specific tuning, thereby reducing information loss.

    % \item 
    \textbf{Improved accuracy in generative recommendation:} Extensive experiments on three real-world datasets -- MIND for news recommendation, CDs for music recommendation, and H\&M for fashion recommendation -- demonstrate that \model{} achieves superior performance compared with RQ-VAE based methods across multiple recommendation scenarios, highlighting its effectiveness.
% \end{itemize}
% 
\section{Related Work}

\subsection{LLMs for Recommendation}

Generally, the emerging techniques of LLMs for enhancing recommender systems can be grouped into three paradigms, namely pre-training, prompting, and fine-tuning~\cite{zhao2024recommender}. 

\paragraph{\textbf{Pre-training.}}~\cite{prec,wu2020ptum,cui2022m6,p5,greenrec} Research in this paradigm typically involves tasks designed to model diverse user behaviors and aims to develop a fundamental recommendation model. For instance, PTUM~\cite{wu2020ptum} employs two pre-training tasks: masked behavior prediction and next K behavior prediction. Similarly, Cui et al.~\cite{cui2022m6} introduce M6, which utilizes an auto-regressive generation task and a text-infilling objective. Additionally, Geng et al.~\cite{p5} propose P5, a model that integrates multiple recommendation tasks within a unified framework to pre-train a foundational recommendation model.

\newcommand{\secondary }[1]{\textcolor{gray}{#1}}

% Please add the following required packages to your document preamble:
% \usepackage{multirow}
\begin{table*}[t]
\centering
\renewcommand{\arraystretch}{1.2} 
\caption{Comparison with previous semantic identifier-based generative recommendation methods and proposed \model{}.}\label{tab:comparison}

\setlength\tabcolsep{4pt}

% \resizebox{\linewidth}{!}{

\begin{tabular}{c|cc|c|cc}
\toprule
\textbf{Method} & \multicolumn{2}{c|}{\textbf{Embedder}} & \textbf{Quantizer} & \multicolumn{2}{c}{\textbf{Recommender}} \\
\midrule
\ & \textbf{Semantic} & \textbf{Collaborative} & \textbf{Model} & \textbf{Model} & \textbf{Alignment Task} \\
\midrule
TIGER~\cite{tiger} & SentenceBERT & N/A & RQ-VAE & Transformer & $\times$ \\
\midrule
LC-Rec~\cite{lc-rec} & Llama1-7B & N/A & RQ-VAE & Llama1-7B & $\checkmark$ \\
\midrule
CoST~\cite{CoST} & SentenceT5 & N/A & RQ-VAE & Transformer & $\times$ \\
\midrule
\secondary{EAGER}~\cite{eager} & \secondary {SentenceT5} & \secondary {DIN} & \secondary  {K-Means} & \secondary {Transformer} & \secondary{$\times$} \\
\midrule
\secondary {LETTER~\cite{letter}} & \secondary {Llama1-7B} & \secondary {SASRec} & \secondary {RQ-VAE} & \secondary {Transformer} & \secondary{$\times$} \\
\midrule
\secondary {TokenRec~\cite{tokenrec}} & \secondary {N/A} & \secondary {LightGCN} & \secondary {MQ-VAE} & \secondary {Llama1-7B} & \secondary{$\checkmark$} \\
\midrule
\model & OPT-350M & N/A & K-Means & OPT-350M & $\checkmark$ \\
\bottomrule
\end{tabular}

\end{table*}

\paragraph{\textbf{Prompting.}}~\cite{xi2023towards,wang2023zero,once} Instead of pre-training an LLM, some studies aim to directly integrate LLMs into the recommendation pipeline without parameter updates, typically through feature augmentation. For instance, Xi et al.~\cite{xi2023towards} propose utilizing LLMs to infer user preferences and factual knowledge about items. Similarly, Wang et al.~\cite{wang2023zero} employ LLMs to model user preferences.

\noindent\paragraph{\textbf{Fine-tuning.}}  This line of research seeks to leverage the capabilities of existing powerful LLMs with fine-tuning. Fine-tuning is a critical step in aligning LLMs with various downstream recommendation tasks. Related studies either adopt full-model fine-tuning~\cite{friedman2023leveraging, shen2023towards} or employ parameter-efficient fine-tuning techniques~\cite{bao2023tallrec, wu2024exploring}, such as LoRA~\cite{lora}, to reduce computational resource requirements.

\subsection{Generative Recommendation}
Generative recommenders bypass filtering or ranking by directly generating recommendations from user interactions or sequential patterns~\cite{li2023large}. Traditional models such as SASRec~\cite{sasrec} and BERT4Rec~\cite{bert4rec}, along with language model–based recommenders like P5~\cite{p5} and VIP5~\cite{vip5}, represent items with unique identifiers and predict the next item by selecting the most probable candidate.

To incorporate item content knowledge, TIGER~\cite{tiger} introduced semantic identifiers that can be shared across items, replacing unique identifiers. This concept has been refined by subsequent semantic tokenization approaches~\cite{lc-rec, jin2023language, uist} (see Table~\ref{tab:comparison}). Notably, methods in gray (EAGER~\cite{eager}, LETTER~\cite{letter}, and TokenRec~\cite{tokenrec}) also integrate collaborative features from simple recommenders into identifiers, though they rely on rich interactions and tend to be unstable in practice.

The resulting semantic identifiers are used in generative recommenders, shifting the training objective from next-item to next-code prediction. This shift constrains the search space for each position, thereby enhancing inference performance.

However, existing methods either depend on differentiable vector quantization -- which is challenging to train -- or on hierarchical clustering schemes (e.g., EAGER’s hierarchical K-Means), which are ill-suited for real-world scenarios where items exhibit multi-aspect information. Given these limitations, this paper reevaluates and refines the standard semantic tokenization framework.
\begin{figure*}[ht]
    \centering
    \includegraphics[width=.9\linewidth]{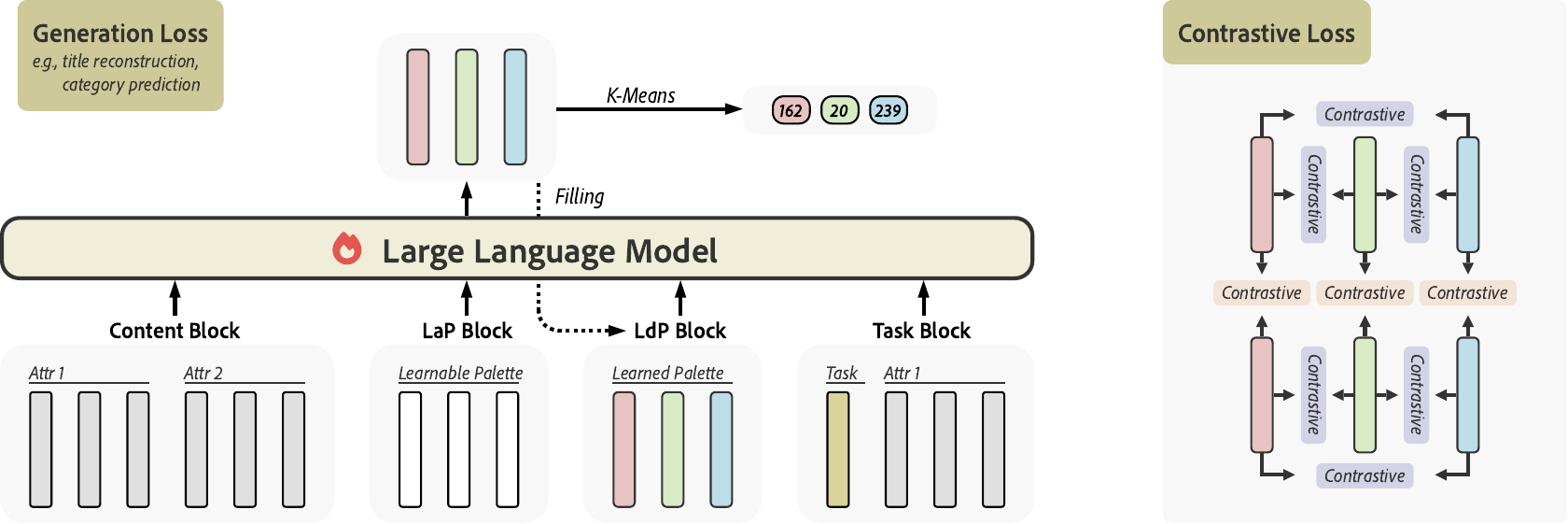}
    \caption{Detailed architecture of \model{}. 
    %Items are initially compressed into embeddings and subsequently clustered to generate semantic tokens, which are then incorporated into the downstream recommender system. Such ``text-to-token'' conversion will be achieved by hierarchical attention masking mechanism and self-supervised tasks.
    For each item, an item palette is learned with a text-based generative (reconstruction or prediction) task and a contrastive learning task. 
    }
    \label{fig:tokenizer}
\end{figure*}

\section{Preliminaries}

Traditional recommender systems use unique item identifiers mapped to learnable vectors (embeddings) to represent items during training, leading to the challenges mentioned earlier. Recent studies~\cite{tiger,lc-rec,uist} have shown that semantic tokenization is becoming an increasingly popular approach for embedding item content into identifiers. In this section, we focus on two key aspects: i) generating semantic identifiers using standard approaches, and ii) utilizing these semantic identifiers in downstream recommendation models.

\subsection{Semantic Tokenization}

Semantic tokenization approaches aims to generate the semantic identifier, a combination of discrete tokens, that can represent items, and can be shared and linked across different items, based on item content or collaborative insights. As illustrated in Table~\ref{fig:comparison}, the standard semantic tokenization pipeline~\cite{tiger,lc-rec} consists of an embedder, a quantizer, and a recommender.

The embedder, typically a pretrained language model such as SentenceBERT~\cite{sentencebert} or LLaMA~\cite{llama}, is often used because textual features are the primary modality in most recommendation scenarios. Leveraging the rich knowledge from pretrained corpora, these embedders extract a robust item embedding $\mathbf{e} \in \mathrm{R}^{d}$ based on the corresponding text content with variable length.

Next, a residual quantizer~\cite{rq-vae} is used to discretize each item embedding $\mathbf{e}$ into structure-aware discrete tokens, using multi-layer ($L$) codebooks, with each codebook containing $K$ code vectors. Each item is (expected to be) mapped to a combination of codes with a length of $L$, where each position is selected from the corresponding codebook. Theoretically, the representation space for $L$-layer codebooks is $K^L$, which means that even a much smaller $K$ and $L$ can effectively represent a total of $N$ items, even when $N \gg K$. Consequently, the substantial memory required for item embeddings in traditional recommenders, i.e., $N \times D$ where $D$ is the embedding dimension, can be compressed into a significantly smaller, logarithmic space, i.e., $K \times L \times D$.

\subsection{Downstream Recommenders}

Once the semantic codes are obtained, they can be integrated into various recommendation models. In sequential recommendation~\cite{tiger,eager}, the user sequence is replaced by a flattened sequence of semantic codes, shifting the task from next-item prediction to next-code prediction. Alternatively, these codes can enhance the recommendation ability of large language models by integrating collaborative knowledge~\cite{lc-rec,tokenrec}.

%\subsection{Limitations}

% However, this standard pipeline has several limitations.

% First, the item content in recommendation scenarios often exhibits a \textbf{data distribution bias} compared to the pretrained language model corpus, as previous works do not apply post-pretraining.

% Second, both the embedder and quantizer contribute to \textbf{information loss}. A single embedding generated by the language model cannot capture the entirety of the original content. Furthermore, the quantization process amplifies this loss, as only a few discrete tokens are used to represent the item embedding.

% \input{Content/Figures/Tokenizer}

% Third, the quantizer, typically a residual quantizer, is \textbf{not easy to train}. Issues like codebook collapse and item collision are common, requiring additional techniques such as K-Means initialization and code replacement to mitigate these problems.

% Fourth, \textbf{the knowledge transfer} between the embedder and the recommender is \textbf{fragmented}, with the semantic identifier serving as the sole connection. No additional support is provided by the embedder to the recommender.
\section{Proposed Approach: \model{}}

The standard semantic tokenization framework frequently utilizes RQ-VAE technology to discretize item embeddings from a singular perspective. Key challenges include training deficiencies such as codebook collapse in differentiable vector quantization and a reliance on item embeddings from pretrained language models that are misaligned with the pertinent recommendation contexts. These limitations necessitate a critical reevaluation of the framework. In this section, we introduce a novel \model{} framework designed to overcome these challenges.

\begin{figure}[t]
    \centering
    \includegraphics[width=\linewidth]{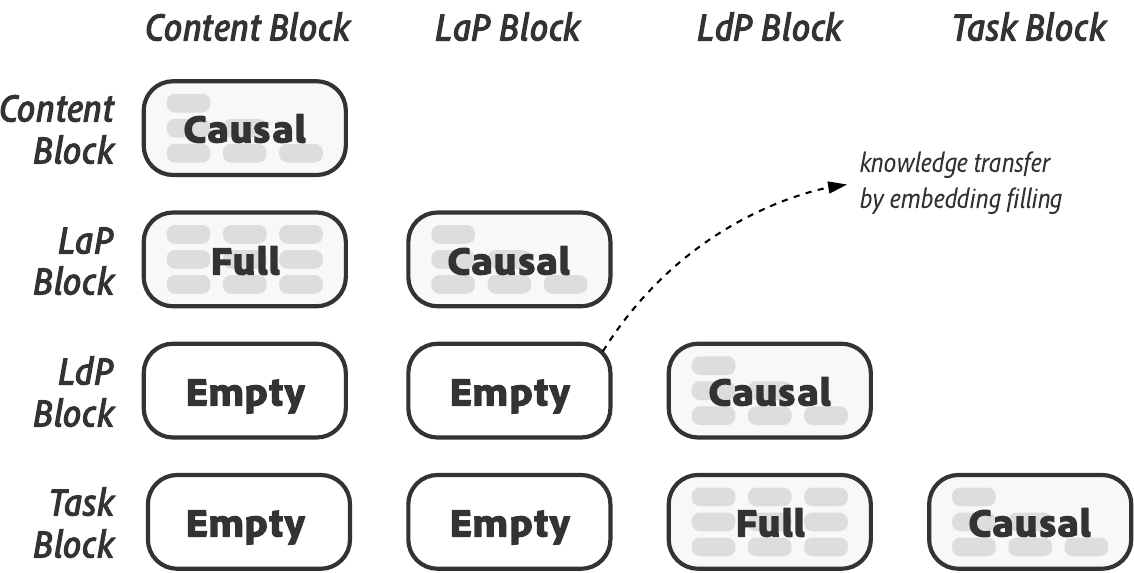}
    \caption{Hierarchical attention masking scheme. The value at the intersection of the $i$-th row and $j$-th column indicates the participation mode of the $j$-th column block in the attention computation for the $i$-th row block. Values where $i = j$ represent the inner-block masking, while those with $i < j$ are defined as inter-block masking.}
    \label{fig:attention}
\end{figure}

% \textbf{Firstly}, our \model{} compresses item content into multi-view item palette, in the form of multiple embeddings, by tuning a large language model on domain-specific data. Compared with using a single embedding to represent each item, using item palette reduce the information loss when reconstructing the original text content. Furthermore, it also reduces the distribution gap between the language model and the recommendation scenario when comparing the previous researches that use a frozen language model for extraction. \textbf{Secondly}, since the \model{} already generate multiple embeddings for each item, residual quantizers are not required to discretize the embedding into discrete tokens: simple, training-free algorithms can be applied to cluster the item palette to directly generate discrete tokens. Such simple solution will avoid from suffering the codebook collapse problem.
\textbf{Initially}, we developed a multi-aspect item palette to store and interpret item content knowledge. This matrix, consisting of several embeddings, compresses variable-length item content through domain-specific text-level reconstruction tasks, allowing the language model to adapt to the pertinent recommendation scenario. Unlike RQ-VAE, which discretizes only a single embedding, our approach enriches the training process with a broader array of reconstructable item content information, thereby minimizing information loss. \textbf{Moreover}, the inherently multiple embeddings of our item palette leverage a training-free clustering algorithm to determine semantic identifiers for each embedding, circumventing the technical challenges posed by differentiable vector quantization. \textbf{Lastly}, our item palette embeddings are independent, mutually exclusive, semantically parallel and equitable, free from hierarchical constraints. This enhanced representational freedom surpasses RQ-VAE and enables the articulation of multi-perspective features of items.

\subsection{Architecture of \model{}}

To enable the compression of variable-length text content into fixed-length item palettes, we introduce the \model{}, compatible with any decoder-only large language model. This model harnesses a block-wise input scheme, hierarchical attention masking, self-supervised generative tasks, and auxiliary contrastive tasks, inspired by the gisting framework~\cite{gist} which condenses prompts into concise tokens to optimize inference in natural language processing.

\noindent\textbf{Block-wise input scheme.} The input sequence is structured into four blocks: content, learnable palette (LaP), learned palette (LdP), and task. For an item $\mathbf{x}$, such as a news article with $m$ attributes ${\mathbf{a}_1, \mathbf{a}_2, \cdots, \mathbf{a}_m}$, we select the first $r \leq m$ attributes for the content block, formulated as:
\begin{equation}
    \texttt{<content>} = [\mathbf{a}_1; \mathbf{a}_2; \cdots, \mathbf{a}_r],
\end{equation}
where $[;]$ signifies concatenation. For instance, a typical article with three ($m=3$) attributes (title, abstract, category) may utilize title and abstract ($r=2$) for the content block. One possible content block can be ``\textttl{Yellowstone tourist injured ...} \textabs{A tourist suffered severe burns ...}''.

The learnable palette block (LaP) consists of $L$ predefined tokens:
\begin{equation}
    \texttt{<LaP>} = [\textdto{}, \textdtt{}, \textdtr{}].
\end{equation}
For the simplicity, $L$ is set to 3 in Figure~\ref{fig:tokenizer}. These tokens possess learnable yet randomly initialized embeddings that facilitate the integration of item content into actionable insights via a transformer network. \textbf{Item Palette is the output embeddings of the LaP.}

A learned palette block (LdP) mirrors the LaP in length but initially contains placeholder tokens:
\begin{equation}
    \texttt{<LdP>} = [\textpho{}, \textpht{}, \textphr{}.
\end{equation}
It will replaced with output embeddings from the LaP before processing. This arrangement supports text-level reconstruction, ensuring the task block leverages the item palette for text generation from the first transformer layer.
% while it will be filled with the output palette embeddings from the LaP before it is fed into the transformer network. This design is necessary for the text-level reconstruction task, ensuring that the task block is guided by the item palette to generate text.

The task block concludes the sequence with a token specific to the task and the answer sequence:
\begin{equation}
    \texttt{<task>} = [t_i; \mathbf{a}_i],
\end{equation}
where $t_i$ denotes a task, varying based on the attribute index, aimed at either reconstructing (when $0 < i \leq r$) the attribute used in the content block or predicting (when $r < i \leq m$) unseen attributes. Two possible examples can be:
\texttt{``\texttsk{<reconstruct\_title>} \textttl{Yellowstone tourist injured ...}''} when $i=1$ and \texttt{``\texttsk{<predict\_category>} \textcat{travel}''} when $i=3$.

\noindent\textbf{Hierarchical Attention Masking.} Decoder-only large language models typically employ causal attention masks, which restrict each token to only attend to preceding tokens and itself, excluding future tokens. This traditional design does not suit scenarios where \textbf{only} the item palette is allowed to influence task output generation. Consequently, we implement a hierarchical attention masking scheme that encompasses both inner-block and inter-block masking.

As depicted in the diagonal of Figure~\ref{fig:attention}, \textit{inner-block masking} maintains causal attention to ensure sequential knowledge retention. In contrast, \textit{inter-block masking} permits configurations of either unrestricted or no attention: the content block fully engages with the learnable palette block, while the learned palette block entirely focuses on the task block. All other inter-block attentions are disabled. 

\subsection{Learning Multi-aspect Item Palette}

We will perform domain-adaptive tuning on a decoder-only large language model to compress variable-length item content knowledge into a multi-aspect item palette. Our training objectives are designed to optimize the item palette’s functionality: First, it should capture and retain ample item content information to facilitate accurate content reconstruction and minimize information loss. Second, each embedding in the item palette should be as independent and mutually exclusive as possible to ensure minimal redundancy. Therefore, we devise the following self-supervised training tasks.

% Notably, we will employ low-rank adaptation (LoRA)~\cite{lora}, a parameter-efficient fine-tuning approach, during the following training on the language model. We will also freeze the pretrained word embeddings while tuning the palette embeddings and task embeddings.

\noindent\textbf{Generative Reconstruction or Prediction.} To fully incorporate content knowledge into the item palette, we create distinct input samples for each item by iterating through task ID $t_i$, ranging from $1$ to $m$. Each task ID corresponds to a unique reconstruction or prediction challenge as discussed before. Utilizing the specialized \model{} architecture, the language model undergoes tuning with the next-token prediction task:
\begin{equation}
    \mathcal{L}_\text{gen} = -\log P(a_{i,j+1} | a_{i,1}, a_{i,2}, \dots, a_{i,j}),
\end{equation}
which is optimized using cross-entropy loss, where $a_{i,j}$ denotes the $j$-th token of the attribute $\mathbf{a}_i$.

To facilitate the transfer of item palette embeddings to the learned palette block, we employ \textit{dual forward propagation}: during the initial forward pass, we capture the item palette embeddings, subsequently populate the LdP block, and deduce the task output based on the populated LdP block.

\noindent\textbf{Contrastive Learning.} \textit{\intrac{Intra-Palette Contrastive Loss.}} Since the palette embeddings of each item at the same order capture specific perspectives of the item content and are intended for subsequent clustering, it is crucial that the embeddings of different orders for the same item are mutually exclusive and contain minimal redundant information. This ensures that the clustering results are independent of each other. To achieve this, we introduce the intra-palette contrastive learning task, which aims to decrease the similarity between palette embeddings of different orders. 
Specifically, we use Hinge Loss to restrict the similarity of the palette embeddings within each sample to remain within a certain threshold $\alpha_\text{intra}$, which can be formulated as:
\begin{gather}
\bar{\mathbf{B}}_{i,j} = \frac{\mathbf{B}_{i,j}}{\|\mathbf{B}_{i,j}\|_2}, \\
\mathcal{L}_\text{intra} = \sum_{i=1}^{B} \sum_{j=1}^{L} \sum_{k=1, k \neq j}^{L} \max(0, \mathbf{s}(\bar{\mathbf{B}}_{i,j}, \bar{\mathbf{B}}_{i,k}) - \alpha_\text{intra})^2,
\end{gather}
where $\mathbf{B} \in \mathrm{R}^{B \times L \times d}$ ($B$ and $L$ represent batch size and the palette size, respectively) is a batch of palette embeddings, $\bar{\mathbf{B}}_{i,j}$ represents the normalized palette embedding of the $j$-th item in the $i$-th order. $\mathbf{s}$ denotes the cosine similarity function.

\textit{\interc{Inter-Palette Contrastive Loss.}} Furthermore, to address the code collision problem -- where multiple item palettes may collapse to the same semantic identifier -- we introduce a contrastive loss between embeddings of item palettes at the same order, encouraging them to remain semantically distinct. Specifically, we treat all other samples within the batch as negative samples for the current sample. We use Hinge Loss to restrict the similarity of the palette embeddings of the same order between negative sample pairs to remain within a certain threshold $\alpha_\text{inter}$, which can be formulated as:
\begin{gather}
\mathcal{L}_\text{inter} = \sum_{i=1}^{B} \sum_{k=1, k \neq i}^{B} \sum_{j=1}^{L} \max(0, \mathbf{s}(\bar{\mathbf{B}}_{i,j}, \bar{\mathbf{B}}_{k,j}) - \alpha_\text{intra})^2.
\end{gather}

Therefore, the final training task of \model{} is optimized by:
\begin{equation}
\mathcal{L}_\text{LAMIA} = \mathcal{L}_{\text{gen}} + \gamma \mathcal{L}_{\text{cl}} = \mathcal{L}_{\text{gen}} + \gamma \left( \mathcal{L}_{\text{intra}} + \mathcal{L}_{\text{inter}}\right),
\end{equation}
where $\gamma$ is the hyperparameter that balances the contribution of the generative and contrastive loss.

% Notably, our approach differs the gisting framework~\cite{gist} in several ways: firstly, our compression targets item content in recommendation scenarios; secondly, we introduce a learned palette block adjacent to the learnable palette block to ensure the task block is guided by the output of the learnable palette block; thirdly, we design multiple self-supervised learning tasks based on the item palette.

% \input{Content/Figures/CodeTree}

\subsection{Quantization Using A Simple Clustering Algorithm}

Unlike standard pipeline that employ complex, hard-to-train differentiable vector quantization techniques to segment item content embeddings into discrete tokens, our dense tokenizer efficiently maps item content features into reconstructable embeddings -- termed item palette -- that encapsulate domain-specific content. Hence, these dense vectors can be discretized into cluster indices using a training-free clustering approach. We first aggregate the output of the item palette embeddings for all items as follows:
\begin{equation}
    \mathbf{E} = \begin{bmatrix}
        \mathdto{e}_{1,1} & \mathdto{e}_{1,2} & \cdots & \mathdto{e}_{1,n} \\
        \mathdtt{e}_{2,1} & \mathdtt{e}_{2,2} & \cdots & \mathdto{e}_{2,n} \\
        \vdots & \vdots & \ddots & \vdots \\
        \mathbf{e}_{v, 1} & \mathbf{e}_{v, 2} & \cdots & \mathbf{e}_{v, n}
    \end{bmatrix},
\end{equation}
where $n$ denotes the number of items, and $\mathbf{e}_{i,j}$ represents the $i$-th item palette embedding of the $j$-th item with $D$ dimensions.
    
Next, we apply the principal component analysis (PCA) technique~\cite{pca} to reduce the high-dimensional $D$ vectors (over 1024 dimensions) to a lower dimension $d$, such as 32. The principal components are denoted by $\mathbf{\hat{e}}_{i,j}$ for each reduced embedding of $\mathbf{e}_{i,j}$.

Finally, we apply a simple, training-free clustering algorithm, such as K-Means~\cite{kmeans}, to each row of the matrix, denoted as $\mathbf{\hat{E}}[i,:] = [\mathbf{\hat{e}}_{i,1}, \mathbf{\hat{e}}_{i,2}, \cdots, \mathbf{\hat{e}}_{i,n}]$. The resulting cluster indices are organized into a matrix:
\begin{equation}
    \mathbf{C} = \begin{bmatrix}
        \palette{$c_{1,1}$}{pc1} & \palette{$c_{1,2}$}{pc1} & \cdots & \palette{$c_{1,n}$}{pc1} \\
        \palette{$c_{2,1}$}{pc2} & \palette{$c_{2,2}$}{pc2} & \cdots & \palette{$c_{2,n}$}{pc2} \\
        \vdots & \vdots & \ddots & \vdots \\
        \palette{$c_{L,1}$}{white} & \palette{$c_{L,1}$}{white} & \cdots & \palette{$c_{L,n}$}{white}
    \end{bmatrix},
\end{equation}
where $1 \leq c_{i,j} \leq k$ represents the cluster indices, and $k$ is the number of clusters. Thus, each item can be represented by $L$ discrete tokens: $\mathbf{c}_j= [c_{1,j}, c_{2,j}, \ldots, c_{v,j}]$.

\begin{figure}[t]
    \centering
    \includegraphics[width=\linewidth]{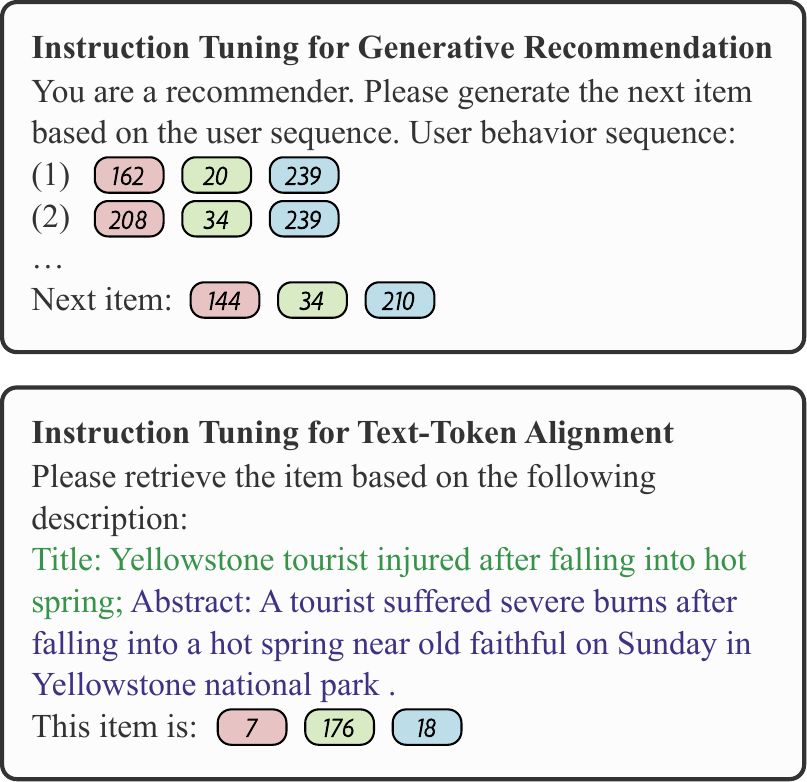}
    \caption{Instruction templates for tuning large language models as generative recommenders. The upper instruction outlines the primary training task of next-item prediction, while the lower instruction details an auxiliary text-token alignment task.}
    \label{fig:recommender}
\end{figure}

\subsection{Generative Recommender}

Like other item tokenization approaches, the item semantic identifiers, or discrete tokens, generated by the clusterer can be utilized in generative retrieval (or sequential recommendation). We can use both the traditional deep learning based recommendation models (DLRMs) or large language models (LLM as RS) for the generative recommendation training and inference. When using large language models as recommenders, we can additionally design the text-token alignment task to enhance the token comprehension capabilities, following LC-Rec~\cite{lc-rec}, as illustrated in Figure~\ref{fig:recommender}.

\noindent\textbf{Training.} 
Given a user behavior sequence in which each item is represented by semantic tokens, this sequence is formalized as:
\begin{equation}
\mathrm{u} = [\palette{$u_{1,1}$}{pc1}, \palette{$u_{1,2}$}{pc2}, \ldots, \palette{$u_{1,v}$}{white}, \palette{$u_{2,1}$}{pc1}, \palette{$u_{2,2}$}{pc2}, \ldots].
\end{equation}

When employing either DLRMs or LLMs as sequential recommendation backbones, the next-item prediction task is always a principal task, which can be roughly formalized as follows:
\begin{equation}
\mathcal{L}_\text{nip} = -\sum_{i=1}^{l} \sum_{j=1}^{L} \log P(u_{i,j+1} | u_{1,k}, \cdots, u_{i-1, k}, u_{i, 1}, \dots, u_{i,j}),
\end{equation}
where $l$ denotes the number of items in the user sequence, and $0 < k \leq L$.

When employing a LLM as the backbone, the user sequence is concatenated using natural language; however, the loss computation exclusively considers semantic tokens, disregarding the natural language components.
Additionally, we employ a text-token alignment task as a supplementary training object for the LLM. Following the instructions presented in Figure~\ref{fig:recommender}, the sequence can be structured as: $\mathrm{s} = [s_1, \cdots, s_l, \palette{$c_1$}{pc1}, \palette{$c_2$}{pc2}, \cdots, \palette{$c_L$}{white}]$. Therefore, the alignment task can be formalized as:
\begin{equation}
\mathcal{L}_\text{align} = -\sum_{i=1}^{v} \log P(c_{i+1} | s_k, \cdots, \palette{$c_1$}{pc1}, \dots, \palette{$c_i$}{white}),
\end{equation}
where $l$ is the length of the natural language, and $0 < k \leq l$ represents any tokens in the language sequence.

\noindent\textbf{Inference.} The recommender will generate the semantic tokens for the next item in an autoregressive manner. Following previous work~\cite{tiger,lc-rec}, beam search~\cite{beamsearch} is applied to maintain top-K token combination.

\begin{table}[]
\centering
\renewcommand{\arraystretch}{1.2} 
\caption{Dataset statistics.}\label{tab:dataset}

\begin{tabular}{l|l|l|l}
\toprule
 & \textbf{MIND} & \textbf{CDs} & \textbf{H\&M} \\
\midrule
\#Items & 25,634 & 19,684 & 15,889 \\
\#Users & 45,000 & 45,000 & 45,000 \\
\#Finetune & 40,000 & 40,000 & 40,000 \\
\#Test & 5,000 & 5,000 & 5,000 \\
Avg. User Length & 11.78 & 5.19 & 8.67 \\
Avg. Item Appearance & 20.69 & 11.70 & 22.44 \\
\bottomrule
\end{tabular}
\end{table}

% Please add the following required packages to your document preamble:
% \usepackage{multirow}
% \begin{table}[]
% \centering
% \renewcommand{\arraystretch}{1.2} 
% \caption{Dataset statistics.}\label{tab:dataset}

% \setlength\tabcolsep{3pt}

% \resizebox{\linewidth}{!}{
% \begin{tabular}{c|c|c|c|c}
% \toprule
%  & \textbf{Content Block} & \textbf{Task Target} & \multicolumn{2}{c}{\textbf{Dataset}} \\
% \midrule
% \multirow{4}{*}{\makecell{MIND \\ ($m$=4, $k$=2)}} 
%  & \multirow{4}{*}{\makecell{(title, abstract)}}  & title & \#Items & 25,634 \\
%  & & abstract & \#Users & 45,000 \\
%  & & category & \#Finetune & 40,000 \\
%  & & subcategory & \#Test & 5,000 \\
% \midrule
% \multirow{4}{*}{\makecell{Yelp \\ ($m$=4, $k$=3)}} 
%  & \multirow{4}{*}{\makecell{(name, city, address)}}  & name & \#Items & 73,380 \\
%  & & city & \#Users & 45,000 \\
%  & & address & \#Finetune & 40,000 \\
%  & & state & \#Test & 5,000 \\
% \midrule
% % \multirow{7}{*}{\makecell{H\&M \\ ($m$=7, $k$=7)}}
% \bottomrule
% \end{tabular}
% }
% \end{table}

\begin{table*}[]
\centering
\small
\renewcommand{\arraystretch}{1.2} 

\caption{Overall performance comparison in retrieval scenarios. \textcolor{red}{An asterisk (*)} denotes a inequitable comparison, as the tokenization method (i.e., EAGER) incorporates both content and behavioral embeddings, unlike other methods that use only content information. We use ``TRM'' to represent train-from-scratch transformer network with causal attention mechanism. We use ``3L'', ``6L'' and ``12L'' to indicate the number of transformer layers. We bold the best scores and underline the second.}\label{tab:big-table}

\setlength\tabcolsep{2pt}

\resizebox{\linewidth}{!}{
\begin{tabular}{c|c|c|ccc|ccc|ccc|ccc|ccc|ccc}
\toprule
\multirow{3}{*}{\makecell{Method}} & \multirow{3}{*}{\makecell{Embedder}} & \multirow{3}{*}{\makecell{Recommender}} & \multicolumn{6}{c|}{\textbf{MIND}} & \multicolumn{6}{c|}{\textbf{CDs}} & \multicolumn{6}{c}{\textbf{H\&M}} \\
\cmidrule(lr){4-9} \cmidrule(lr){10-15} \cmidrule(lr){16-21}
 & &  & \multicolumn{3}{c|}{\textbf{Recall \%}} & \multicolumn{3}{c|}{\textbf{NDCG \%}} & \multicolumn{3}{c|}{\textbf{Recall \%}} & \multicolumn{3}{c|}{\textbf{NDCG \%}} & \multicolumn{3}{c|}{\textbf{Recall \%}} & \multicolumn{3}{c}{\textbf{NDCG \%}} \\
\cmidrule(lr){4-9} \cmidrule(lr){10-15} \cmidrule(lr){16-21}
 & &  & 5 & 10 & 20 & 5 & 10 & 20 & 5 & 10 & 20 & 5 & 10 & 20 & 5 & 10 & 20 & 5 & 10 & 20 \\ 
\midrule[1pt]
\multicolumn{21}{l}{\cellcolor{bglight} Item Representation: \textbf{Unique ID}} \\
\midrule[1pt] 
GRU4Rec~\cite{gru4rec} & \multirow{6}{*}{N/A}
& \multirow{6}{*}{N/A} 
& 0.48 & 0.52 & 0.60 & 0.38 & 0.40 & 0.42 
& 0.22 & 0.22 & 0.24 & 0.15 & 0.15 & 0.16 
& 2.62 & 2.80 & 3.28 & 2.24 & 2.37 & 2.60 \\
Caser~\cite{caser} & & 
& 0.88 & 1.02 & 1.28 & 0.73 & 0.78 & 0.86 
& 0.14 & 0.14 & 0.15 & 0.14 & 0.14 & 0.14 
& 2.02 & 2.24 & 2.69 & 1.75 & 1.88 & 1.90 \\
Bert4Rec~\cite{bert4rec} & & 
& 0.62 & 0.74 & 0.80 & 0.32 & 0.32 & 0.34 
& 0.18 & 0.20 & 0.21 & 0.13 & 0.14 & 0.14
& 1.98 & 2.16 & 2.44 & 1.60 & 1.74 & 1.92 \\
SASRec$_\text{3L}$~\cite{sasrec} & & 
& 1.00 & 1.04 & 1.14 & 0.81 & 0.82 & 0.86 
& 0.24 & 0.24 & 0.24 & 0.16 & 0.16 & 0.16 
& 2.24 & 2.60 & 2.98 & 1.97 & 2.09 & 2.18 \\
SASRec$_\text{6L}$ & & 
& 1.08 & 1.26 & 1.56 & 0.91 & 0.97 & 1.05
& 0.10 & 0.18 & 0.18 & 0.09 & 0.12 & 0.12  
& 3.12 & 3.66 & 4.26 & 2.80 & 2.97 & 3.12 \\
SASRec$_\text{12L}$ & & 
& 3.00 & 3.54 & 4.18 & 2.24 & 2.41 & 2.57
& 10.82 & 11.24 & 11.68 & 10.27 & 10.41 & 10.52
& \textbf{14.00} & \textbf{14.36} & \textbf{14.74} & \textbf{12.23} & \textbf{12.35} & \textbf{12.45} \\
\midrule[1pt]
\multicolumn{21}{l}{\cellcolor{bglight} Item Representation: \textbf{Hierarchical Semantic ID}, using \textbf{RQ-VAE} or \textbf{Hierarchical K-Means}} \\
\midrule[1pt]
CoST~\cite{CoST} & SentenceT5 & \multirow{2}{*}{TRM$_\text{3L}$}
& 2.72 & 3.22 & 4.50 & 1.97 & 2.13 & 2.45 
& 1.42 & 1.66 & 1.74 & 3.36 & 4.49 & 5.71 
& 1.32 & 1.58 & 1.92 & 2.80 & 3.75 & 4.76 \\
EAGER~\cite{eager} & SentenceT5\textcolor{red}{*} &  
& 2.00 & 3.48 & 5.32 & 1.36 & 1.83 & 2.51
& 1.56 & 1.74 & 1.92 & 3.71 & 4.60 & 5.89
& 1.28 & 1.54 & 1.88 & 1.10 & 1.33 & 1.52 \\
% LC-Rec & 
% & 1.66 & 2.44 & 3.52 & 1.26 & 1.51 & 1.78
% & 0.00 & 0.00 & 0.00 & 0.00 & 0.00 & 0.00 
% & 1.14 & 1.46 & 1.83 & 0.88 & 1.10 & 1.47 \\
% \model{}$_\textit{(ours)}$ & 
% & 2.08 & 3.02 & 3.56 & 1.50 & 1.79 & 2.03
% & 0.00 & 0.00 & 0.00 & 0.00 & 0.00 & 0.00 
% & 1.28 & 1.76 & 2.20 & 0.82 & 0.98 & 1.09 \\
\midrule
\multirow{4}{*}{TIGER~\cite{tiger}} & \multirow{4}{*}{SentenceBert} & TRM$_\text{3L}$ 
& 2.98 & 4.64 & 6.52 & 2.15 & 2.66 & 3.13 
& 0.00 & 0.00 & 0.00 & 0.00 & 0.00 & 0.00 
& 1.04 & 1.32 & 1.66 & 0.75 & 0.84 & 0.93 \\
 & & \multirow{1}{*}{TRM$_\text{6L}$}
& 2.94 & 3.42 & 4.44 & 2.09 & 2.25 & 2.50 
& 1.72 & 1.88 & 2.20 & 1.34 & 1.40 & 1.48
& 1.14 & 1.42 & 1.72 & 0.88 & 0.97 & 1.04 \\
 & & \multirow{1}{*}{TRM$_\text{12L}$}
& 2.52 & 3.30 & 4.82 & 1.76 & 2.01 & 2.39 
& 1.74 & 2.14 & 2.26 & 1.36 & 1.49 & 1.52 
& 2.00 & 2.46 & 3.12 & 1.51 & 1.65 & 1.83 \\
 &  & BERT$_\text{base}$
& \underline{8.42} & \underline{8.84} & \underline{9.08} & \underline{6.41} & \underline{6.54} & \underline{6.61} 
& \underline{37.00} & \underline{37.84} & \underline{38.32} & \underline{35.11} & \underline{35.39} & \underline{35.51} 
& 5.42 & 6.66 & 7.88 & 4.27 & 4.67 & 4.98 \\
\midrule
LC-Rec~\cite{lc-rec} & Llama-1$_\text{7B}$ & BERT$_\text{base}$
& 6.37 & 6.68 & 6.95 & 4.80 & 4.94 & 5.08 
& 28.62 & 30.24 & 31.13 & 25.80 & 26.10 & 26.34
& 9.34 & 10.75 & 12.08 & 7.84 & 8.26 & 8.62 \\
\midrule[1pt]
\multicolumn{21}{l}{\cellcolor{bglight} Item Representation: \textbf{Parallel Semantic ID}, using \textbf{\model{}} \textit{(ours)}} \\
\midrule[1pt]
\model{} \textit{(ours)} & OPT$_\text{base}$ & BERT$_\text{base}$
& \textbf{9.08} & \textbf{9.96} & \textbf{10.56} & \textbf{7.42} & \textbf{7.71} & \textbf{7.87} 
& \textbf{38.88} & \textbf{39.70} & \textbf{40.04} & \textbf{36.16} & \textbf{36.43} & \textbf{36.52}
& \underline{11.50} & \underline{12.68} & \underline{13.44} & \underline{9.17} & \underline{9.56} & \underline{9.75} \\
\bottomrule
\end{tabular}
}
\end{table*}

\section{Experiments}

\subsection{Experimental Setup}

\textbf{Datasets.}
We conduct experiments on three real-world content-based recommendation dataset, i.e., MIND (news), Amazon CDs (music), and H\&M (fashion). The dataset statistics are summarized in Table~\ref{tab:dataset}.

\textbf{Baseline models.}
We benchmark our proposed \model{} framework against unique ID-based recommenders (GRU4Rec~\cite{gru4rec}, Caser~\cite{caser}, SASRec~\cite{sasrec}, Bert4Rec~\cite{bert4rec}, and P5~\cite{p5}) and semantic code-based recommenders (TIGER~\cite{tiger} and LC-Rec~\cite{lc-rec}, CoST~\cite{CoST}, and EAGER~\cite{eager}). It is important to note that comparing other methods with EAGER is not equitable, as EAGER learns semantic identifiers based on both item content and behavioral knowledge. All code-based recommenders, i.e., TIGER, LC-Rec, CoST, and our \model{}, represent each item using four codes ($v=4$), with a fixed code vocabulary of 256 at each position. 

\textbf{Evaluation Metrics.} We follow the common practice~\cite{tiger,tokenrec} to evaluate the effectiveness of sequential recommenders with the widely used metrics, i.e., Recall and NDCG~\cite{ndcg}. In this work, we use Recall@1, Recall@5, Recall@10, Recall@20, NDCG@1, NDCG@5, NDCG@10, and NDCG@20 for evaluation.
\begin{table*}[]
\centering
\renewcommand{\arraystretch}{1.2} 
\setlength\tabcolsep{2pt}
\caption{Ablation studies. $\mathcal{L}_{cl}$ and $\mathcal{L}_{align}$ represent the use of the contrastive task and alignment task respectively. ``N/A'' denotes that the setting is not applicable.
}\label{tab:ablation}

\resizebox{\linewidth}{!}{
\begin{tabular}{cccc|cccccc|cccccc}
\toprule
\multirow{2}{*}{Quantizer} & \multirow{2}{*}{Recommender} & \multirow{2}{*}{$\mathcal{L}_\text{cl}$} & \multirow{2}{*}{$\mathcal{L}_\text{align}$} & \multicolumn{6}{c|}{\textbf{MIND}} & \multicolumn{6}{c}{\textbf{H\&M}} \\
\cmidrule{5-16}
& & & & R@5 & R@10 & R@20 & N@5 & N@10 & N@20 & R@5 & R@10 & R@20 & N@5 & N@10 & N@20 \\
\midrule
RQ-VAE & BERT$_\text{base}$ & N/A & $\times$ 
& 6.74 & 6.90 & 7.17 & 5.02 & 5.35 & 5.58 
& 5.25 & 6.38 & 6.86 & 3.94 & 4.15 & 4.50 \\
% \model{} & TRM$_\text{3L}$ & $\checkmark$ & N/A
% & 2.08 & 3.02 & 3.56 & 1.50 & 1.79 & 2.03
% & 1.28 & 1.76 & 2.20 & 0.82 & 0.98 & 1.09 \\
\model{} & BERT$_\text{base}$ & $\times$ & $\checkmark$
& 4.30 & 6.06 & 7.94 & 3.41 & 3.57 & 3.83 
& 8.38 & 9.42 & 10.14 & 6.02 & 6.70 & 7.15 \\
\model{} & BERT$_\text{base}$ & $\checkmark$ & $\times$ 
& 8.84 & 9.79 & 10.26 & 7.06 & 7.28 & 7.38 
& 10.88 & 11.80 & 12.64 & 8.55 & 8.93 & 9.08 \\
\model{} & BERT$_\text{base}$ & $\checkmark$ & $\checkmark$ 
& \textbf{9.08} & \textbf{9.96} & \textbf{10.56} & \textbf{7.42} & \textbf{7.71} & \textbf{7.87} & \textbf{11.50} & \textbf{12.68} & \textbf{13.44} & \textbf{9.17} & \textbf{9.56} & \textbf{9.75} \\
\bottomrule
\end{tabular}
}
\end{table*}

\textbf{Implementation Details.} \textit{i) \model{}.} We utilize the pretrained OPT-350M~\cite{opt} as the backbone LLM to learn item palette. Optimization is performed using the Adam~\cite{kingma2014adam} optimizer with a learning rate of 1e-4, a batch size of 128, a LoRA rank of 128, a palette size $L$ of 4, intra-/inter-palette contrastive margins $\alpha_\text{intra}$ and $\alpha_\text{inter}$ of 0.1 and 0.25, palette contrastive weights $\gamma$ of 0.1. \textit{ii) Self-supervised generative tasks.} For the MIND dataset ($m=4$, $k=2$), we use news title and abstract to form the content block, and four generative tasks are designed: generating title, abstract, category and subcategory for the MIND dataset. For the H\&M dataset ($m=k=7$), we use fashion description, product type, product group, appearance name, color master name, color value name, and index name to form the content block, and each attribute will correspond to a generation task. \textit{iii) Clusterer.} We apply PCA~\cite{pca} to reduce 1024-dimensional item embeddings to 64 components, subsequently clustering each position into 256 groups. We set the similarity threshold of the Hinge Loss in Eq. (8) to 0.25 and the loss weight balancing hyperparameter $\gamma$ to 0.1. \textit{iv) Item Collision.} We use an additional index token to ensure items will be mapped into different semantic identifier. \textit{v) Generative recommender.} We set the maximum length of user history sequence to 20 and we use the last item in the sequence as the prediction target. We use the same pretrained OPT-base as the backbone with a learning rate of 5e-4, a batch size of 64 and a LoRA rank of 128. For LLM based recommender, the training starts with the joint learning of generative recommendation task and text-token alignment task. After model convergence, the model will further be tuned by the single generative recommendation task. Early stopping mechanism is used with patience of 5. All the experiments are conducted on a single NVIDIA A100 device with 80GB memory. We release all our code and data for other researches to reproduce our work. We employ the RecBench benchmark~\cite{recbench} for evaluating the recommendation abilities of large language models, for most of our experiments.

\subsection{
%Generative Recommenders for Retrieval
Main Results}

\begin{figure}[t]
\centering
\resizebox{.9\linewidth}{!}{
\begin{tikzpicture}
\begin{axis}[
    xlabel={Semantic Identifier Length},
    ylabel={R@5},
    xmin=2, xmax=9,
    ymin=10, ymax=30,
    xtick={2,4,6,8},
    ytick={10,15,20,25,30},
    legend pos=north west,
    grid=both,
    grid style={line width=.1pt, draw=gray!30},
    major grid style={line width=.2pt, draw=gray!50},
    minor grid style={line width=.1pt, draw=gray!20},
    axis background/.style={fill=gray!5}, % 轻微灰色背景，提升现代感
    every axis label/.append style={font=\small, color=black},
    every tick label/.append style={font=\footnotesize, color=black},
    axis line style={draw=black!80},
    tick align=outside,
    tick pos=left,
    width=0.6\textwidth,
    height=0.45\textwidth,
]
% PaletteNet 曲线
\addplot[
    color=cyan!70!black,      % 青绿色调
    mark=square*,            % 实心正方形标记
    mark options={scale=1.2},
    line width=1.5pt,
    smooth,                  % 平滑曲线
]
coordinates {
    (3,18.4)(4,20.8)(6,25.2)(8,25.8)
};
\addlegendentry{\model{}}

% RQ-VAE 曲线
\addplot[
    color=magenta!70!black,   % 橙色调
    mark=triangle*,          % 实心三角形标记
    mark options={scale=1.2},
    line width=1.5pt,
    smooth,                  % 平滑曲线
]
coordinates {
    (3,16.6)(4,16.8)(6,18.6)(8,15.4)
};
\addlegendentry{RQ-VAE}

\end{axis}
\end{tikzpicture}
}
\caption{Performance comparison of \model{} and RQ-VAE based on different semantic identifier lengths. Experiments are conducted on the MIND dataset. The RQ-VAE curve is produced by the TIGER model.}
\label{fig:length}
\end{figure}

\begin{figure*}[t]
\centering
\resizebox{\linewidth}{!}{
\begin{tabular}{cccc}
  \begin{subfigure}[b]{.3\linewidth}
    \centering
    \includegraphics[width=\linewidth]{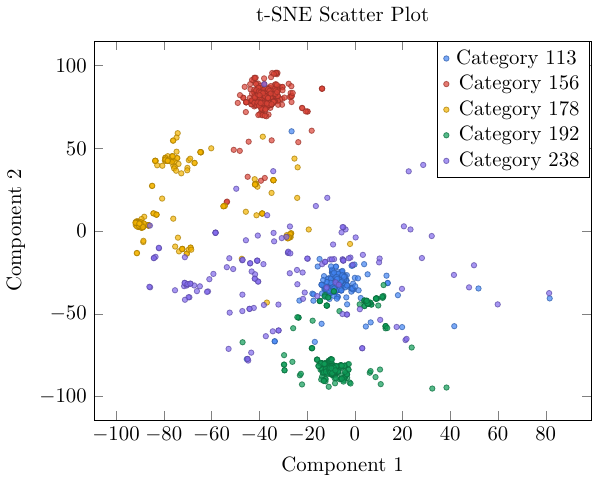}
    \caption{RQ-VAE, first token.}
    \label{fig:rq-1}
  \end{subfigure}
&
  \begin{subfigure}[b]{.3\linewidth}
    \centering
    \includegraphics[width=\linewidth]{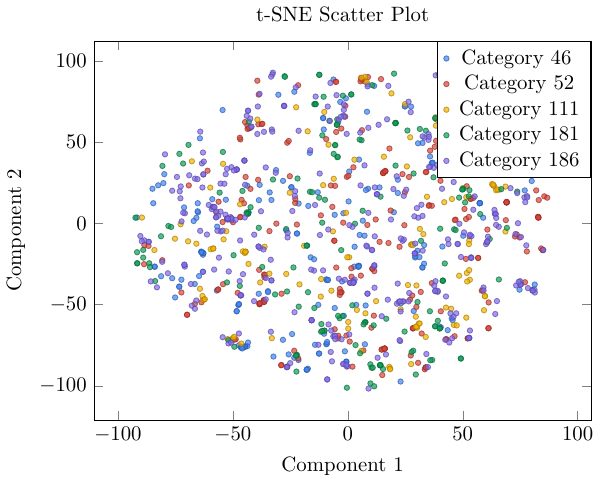}
    \caption{RQ-VAE, second token.}
    \label{fig:rq-2}
  \end{subfigure}
&
  \begin{subfigure}[b]{.3\linewidth}
    \centering
    \includegraphics[width=\linewidth]{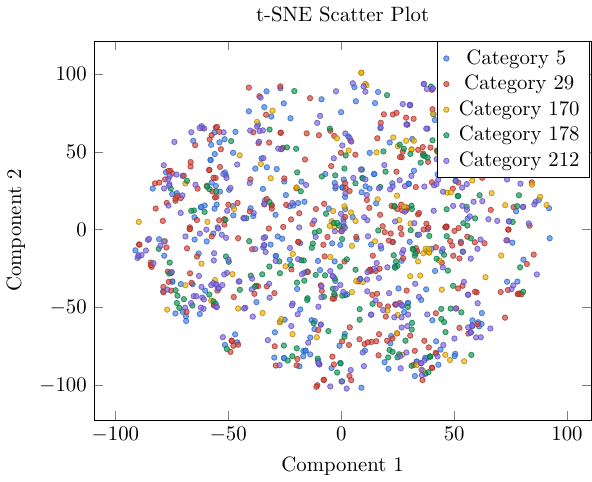}
    \caption{RQ-VAE, third token.}
    \label{fig:rq-3}
  \end{subfigure}
&
\begin{subfigure}[b]{.3\linewidth}
    \centering
    \includegraphics[width=\linewidth]{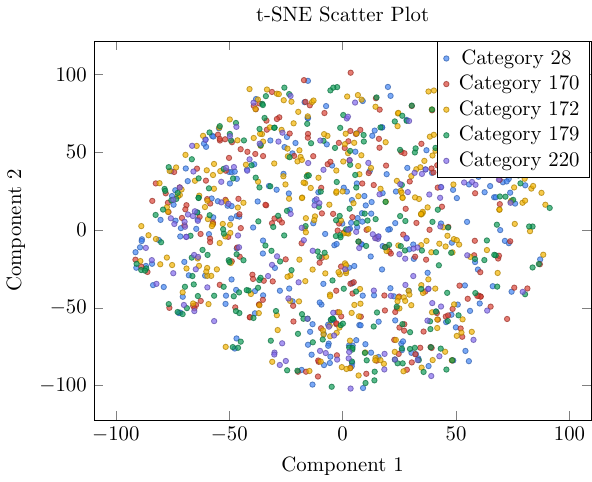}
    \caption{RQ-VAE, fourth token.}
    \label{fig:rq-4}
  \end{subfigure}
\\

  \begin{subfigure}[b]{.3\linewidth}
    \centering
    \includegraphics[width=\linewidth]{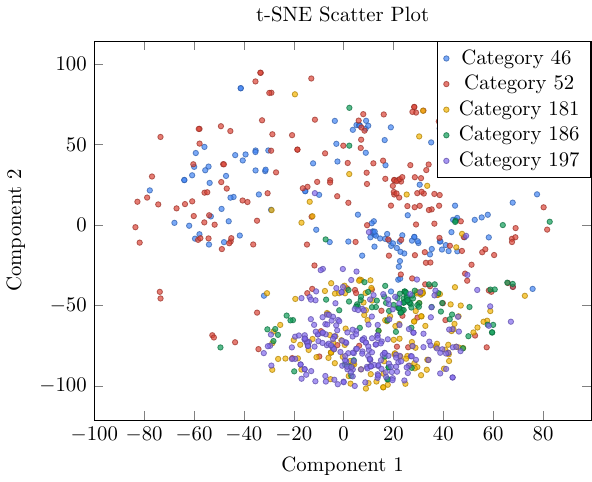}
    \caption{\model{}, first token.}
    \label{fig:our-1}
  \end{subfigure}
&
  \begin{subfigure}[b]{.3\linewidth}
    \centering
    \includegraphics[width=\linewidth]{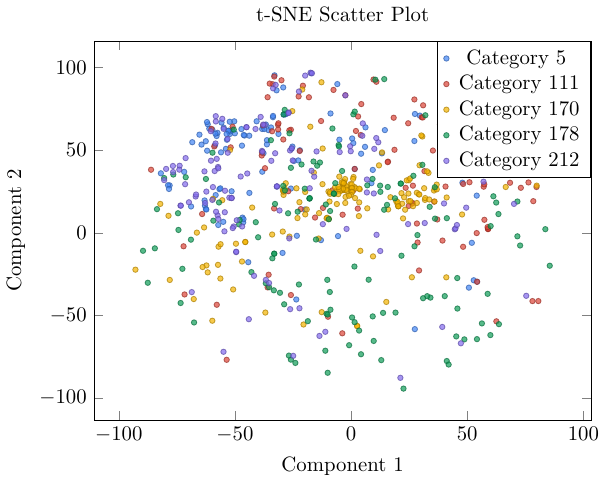}
    \caption{\model{}, second token.}
    \label{fig:our-2}
  \end{subfigure}
& 
\begin{subfigure}[b]{.3\linewidth}
    \centering
    \includegraphics[width=\linewidth]{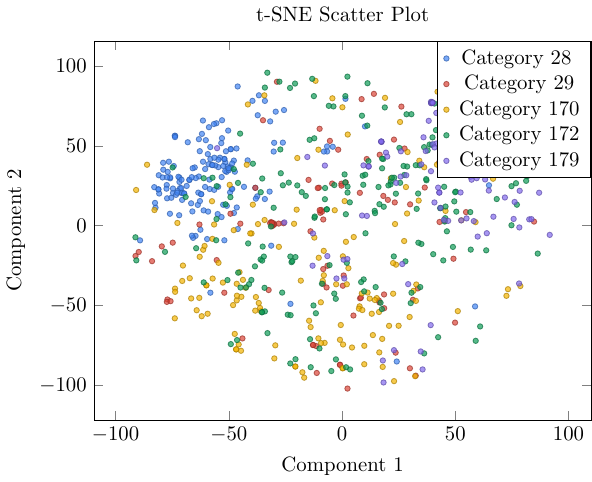}
    \caption{\model{}, third token.}
    \label{fig:our-3}
\end{subfigure}
&
\begin{subfigure}[b]{.3\linewidth}
    \centering
    \includegraphics[width=\linewidth]{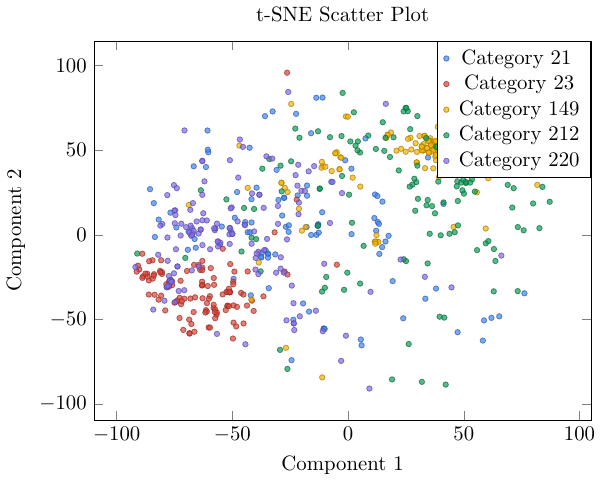}
    \caption{\model{}, fourth token.}
    \label{fig:our-4}
\end{subfigure} \\
\end{tabular}
}
\caption{Visualization for the semantic identifiers produced by RQ-VAE (TIGER~\cite{tiger}) and our \model{} on the MIND dataset. Each point represents an item, with points of the same color indicating items assigned the same automatically generated category label. Spatial proximity between points reflects a close semantic relationship as defined by ground-truth information within each subfigure. Categories are \textbf{RANDOMLY} selected. Our \model{} is designed to generate multi-aspect semantic tokens, with each token at different positions capturing a distinct aspect of item semantics. This design is reflected in the clustering behavior observed within certain categories.}
\label{fig:t-SNE}
\end{figure*}

Table~\ref{tab:big-table} provides an overview of the performance across 11 baselines over two datasets. Drawing from the results, we can derive the following observations: 

\textbf{Firstly}, within the SASRec series, performance generally improves with an increasing number of attention layers, reflecting the scaling behavior of conventional sequential recommenders. Semantic identifier-based methods outperform unique identifier-based ones on the MIND and CDs datasets. However, in the H\&M dataset, the latter performs better -- likely due to low-quality content features, where similar item descriptions (e.g., clothing) correspond to different labels (e.g., color, type), making effective tokenization challenging. 

\textbf{Secondly}, for the TIGER-Transformer series, increasing the number of transformer layers generally leads to improved performance. Notably, replacing TRM$_\text{12L}$ with a 12-layer BERT$_\text{base}$ as the generative recommender results in a significant performance boost. This highlights the value of pretrained language models in capturing rich semantic information and user intent, even when such capabilities are implicitly encoded in semantic IDs. 

\textbf{Thirdly}, LC-Rec, which extracts item content embeddings using LLaMA-1$_\text{7B}$, underperforms compared to TIGER’s SentenceBERT-based model. Although LLaMA-1 possesses broader world knowledge, SentenceBERT--specifically pretrained for sentence representation--achieves better results across all three datasets. This suggests that BERT’s domain knowledge is sufficient for content understanding in this context. 

\textbf{Fourthly}, our proposed \model{} with parallel semantic IDs outperforms the traditional hierarchical semantic ID approach, achieving the best results on two datasets and competitive performance on H\&M, second only to SASRec$_\text{12L}$. These results demonstrate the effectiveness of our model design.

\subsection{Ablation Studies}

Here, we study the effectiveness of various components within our framework. Based on the results from Table~\ref{tab:ablation}, we can make the following observations: 

\textbf{Firstly}, the first row (RQ-VAE-BERT$_\text{base}$) adheres to the conventional RQ-VAE semantic tokenization pipeline, utilizing the same pretrained OPT-base model as our \model{}. Our \model{} outperforms this variant, showcasing the advantages of domain-adaptive tuning and the incorporation of a multi-aspect item palette. 

% \textbf{Secondly}, \model{} with TRM$_\text{3L}$ as the recommender achieves the weakest performance, underscoring the importance of a strong LLM-based backbone in generative recommendation. 

\textbf{Secondly}, \model{} without $\mathcal{L}_{cl}$ lacks the auxiliary contrastive task that is designed to augment the separability of item palette embeddings. Consequently, this configuration may lead to redundant storage of similar item content features within different embeddings, thereby complicating the clustering process. The suboptimal performance of this model variant relative to our fully implemented \model{} underscores the efficacy of incorporating a palette contrastive task. 

\textbf{Thirdly}, \model{} without $\mathcal{L}_{align}$, trained solely with the sequential recommendation task (i.e., next-item prediction), performs less effectively compared to our \model{}. Introducing the text-token alignment task is crucial for bridging the gap between text and tokens, forcing the LLMs to learn the real semantics behind the discrete tokens (semantic identifiers), thereby enhancing the semantic understanding of user sequences. 

\subsection{Effect of Semantic Identifier Length}

We also explore the effect  of semantic identifier length (or item palette) on performance, as illustrated in Figure~\ref{fig:length}. The analysis reveals that for both RQ-VAE and our \model{}, performance improves as the identifier length increases up to 6. Beyond this point, performance for RQ-VAE declines, whereas \model{} continues to show slight improvements when the length reaches 8. This divergence in performance is attributed to the hierarchical and residual discretization method employed by RQ-VAE, where later discrete tokens tend to contain less information than earlier ones, diminishing their utility and adversely affecting overall performance. Conversely, our \model{} adopts a multi-aspect item palette approach, treating each palette embedding equitably, with each embedding encapsulating a segment of the item’s content. Therefore, increasing the length of the palette enhances the granularity and richness of the representation.

\subsection{Visualization}

Finally, we visualize semantic identifiers from both RQ-VAE and our \model{} using the t-SNE~\cite{tsne} technique to project item content embeddings extracted by SentenceT5~\cite{sentenceT5} into two dimensions. Shorter distances between points in subfigures suggest closer semantic relationships between items. Each semantic token in our study has 256 potential categories, from which we randomly selected five for visualization in Figure~\ref{fig:t-SNE}, revealing distinct clustering behaviors. \textbf{Firstly}, RQ-VAE’s initial token clustering reflects its use of hierarchical techniques that group items based on their original embeddings, but residual influences cause dispersion in subsequent layers.  \textbf{Secondly}, our \model{} displays focal clusters for each token (e.g., Category 197 in Figure~\ref{fig:our-1}, Category 5 in Figure~\ref{fig:our-2}), illustrating that primary item semantics, as deduced by t-SNE, are represented across different tokens, each corresponding to distinct aspects, necessitating diverse token positions for comprehensive semantic learning.

\section{Conclusion}

In this paper, we introduce \model{}, a novel semantic tokenization framework designed for generative recommendation. Unlike existing methods that primarily rely on pretrained embeddings and RQ-VAE techniques, our \model{} framework focuses on learning a multi-aspect item palette through text-based reconstruction. This approach not only generates more diverse semantic tokens but also avoids the training instabilities associated with RQ-VAE. Our experimental results demonstrate the effectiveness of \model{}, showing significant improvements in generative recommendation metrics.

While \model{} provides a new perspective for learning parallel and comprehensive semantic tokens, the self-supervised fine-tuning of large language models using a text-level reconstruction task can be time-consuming. In the future, we plan to explore strategies to expedite the multi-aspect palette learning process.

\section*{GenAI Usage Disclosure}
ChatGPT was utilized during the preparation of this paper to assist with text refinement and proofreading, enhancing the clarity and coherence of the writing. The content and research of this paper were developed independently, without the use of GenAI tools.

%ChatGPT was employed during the preparation of this paper to assist with various tasks, including text refinement, grammar checking, proofreading, and overall enhancement of the clarity and coherence of the writing. The tool helped improve sentence structure, correct grammatical errors, and ensure that the language used was clear and concise. While the primary content and research were developed independently, the use of ChatGPT significantly contributed to the final quality of the manuscript.

%%
%% The next two lines define the bibliography style to be used, and
%% the bibliography file.
\balance
\bibliographystyle{ACM-Reference-Format}
\bibliography{Lamia}

\appendix

\end{document}